\documentclass[amsmath,amssymb,twocolumn,groupedaddress]{revtex4-1}

\usepackage{graphicx}
\usepackage{epstopdf}
\usepackage{pstricks}
\usepackage{epsf}
\usepackage{natbib}
\usepackage{eqnarray,amsmath}
\newcommand{\be}{\begin{equation}}\newcommand{\ee}{\end{equation}}
\newcommand{\ba}{\begin{array}{l}}\newcommand{\ea}{\end{array}}
\newcommand{\baa}{\begin{eqnarray}}\newcommand{\eaa}{\end{eqnarray}}
\newcommand{\lab}[1]{\label{#1}}\newcommand{\re}[1]{(\ref{#1})}
\newcommand{\ci}[1]{\cite{#1}}

\DeclareMathOperator{\sn}{sn}
\DeclareMathOperator{\dn}{dn}
\DeclareMathOperator{\cn}{cn}
\DeclareMathOperator{\am}{am}

\begin{document}
\title{The stationary sine-Gordon equation on metric graphs:\\ Exact analytical solutions for simple topologies}
\author{K. Sabirov$^a$, S. Rakhmanov$^b$, D. Matrasulov$^b$, H. Susanto$^c$}

\affiliation{
$^a$National University of Uzbekistan, Vuzgorodok, Tashkent 100174,Uzbekistan\\
$^b$Turin Polytechnic University in Tashkent, 17 Niyazov Str.,
100095,  Tashkent, Uzbekistan\\
$^c$Department of Mathematical Sciences, University of Essex, Wivenhoe
Park, Colchester CO4 3SQ, UK}

\begin{abstract}
 We consider the stationary sine-Gordon equation on metric graphs  with
simple topologies. The vertex boundary conditions are provided by flux conservation and matching of derivatives at the star graph vertex. Exact analytical solutions are obtained. It is shown that the method can be extended for tree and other simple graph topologies. Applications of the obtained results to branched planar Josephson junctions and Josephson junctions with tricrystal boundaries  are discussed.

\end{abstract}

 \maketitle

\section{Introduction}

Nonlinear wave equations have found numerous applications in
different topics of physics and natural sciences (see, e.g.,
\ci{ablowitz,Kivshar,scott,Kivshar1,dauxois,Panos}). Recently they
have attracted much attention in the context of soliton transport
in networks and branched structures
\ci{zar2010,Zarif2011,adami2011,adami-eur,adami-jpa,adami2013,Karim2013,noja,Noja2015,UGSBM15,Hadi2005,caputo14}.
Wave dynamics in networks can be modeled by nonlinear evolution equations on metric graphs.
This fact greatly facilitates the study of soliton transports in branched systems.
Metric graph is a system of bonds which are assigned a length and connected at the
vertices according to a rule, called "topology of a graph". Solitons and other  nonlinear waves in  branched systems appear in different systems of
 condensed matter, polymers, optics, neuroscience, DNA and many other systems.
In condensed matter very important branched systems, where
solitons can appear are the Josephson junction networks
\ci{Ovchinnik}-\ci{Luca1}. The phase difference in a Josephson
junction obeys sine-Gordon equation \ci{Baron}. Josephson junction networks can therefore be effectively modelled by the sine-Gordon equation on metric graphs. The early treatment of superconductor networks consisting of  Josephson junctions meeting
at one point dated back to \ci{Nakajima,Nakajima1}. An
interesting realization of  Josephson junction networks at
tricrystal boundaries  was discussed earlier in \ci{Kogan2000}, which inspired later detailed study of the problem using the sine-Gordon equation on networks in \ci{Hadi,Hadi2004,Hadi2005}.
Discrete sine-Gordon equations were also used in \ci{sodano-epl09,Ovchinnik,Luca1}
to describe different networks of Josepshon junctions having several junctions on each wire of a network.
Recently, a 2D sine-Gordon equation on networks was studied by considering $Y$ and $T$ junctions \ci{caputo14}. Discrete sine-Gordon equations on networks were also considered in \ci{caputo15}.

In this paper we address the problem of stationary sine-Gordon
equations on metric graphs by focusing on exact analytical
solutions for simple graph topologies. Such a one-dimensional, stationary sine-Gordon equation describes, for instance, the transverse component of the phase difference in a 2D Josephson junction in a constant magnetic field. The derivative of the phase difference presents the local magnetic field in the system \ci{Kupl1999,Kupl2006,Kupl2007}.

Planar Josephson junctions were studied in \ci{Kupl2006,Kupl2007} on the basis of  solutions of the stationary sine-Gordon equation on a finite interval.  Here, we use a similar approach to solve  the stationary sine-Gordon equation on metric graphs. The vertex boundary conditions providing connection of the graph
bonds at the branching points are derived from the flux conservation and
continuity of the weights of wavefunction derivatives. 
The model proposed in this work can be used to describe static solitons in 2D Josephson junctions interacting with constant magnetic field  \ci{Kupl2006,Kupl2007}. The results are then extended for metric tree graphs consisting of finite bonds. The study can be generalized for other simple graph topologies which can be constructed using star and loop graphs.

This paper is organized as follows. In the next section we give a formulation of
the problem together with the boundary conditions for the static sine-Gordon
equation on a star graph. Section 3 presents the derivation of the exact analytical solutions for different special cases. In section 4, we extends the treatment for metric tree graphs. In section 5, we explore the stability of the obtained solutions. Finally, Section 6 presents some concluding remarks.

\begin{figure}[ht!]
\includegraphics[width=80mm]{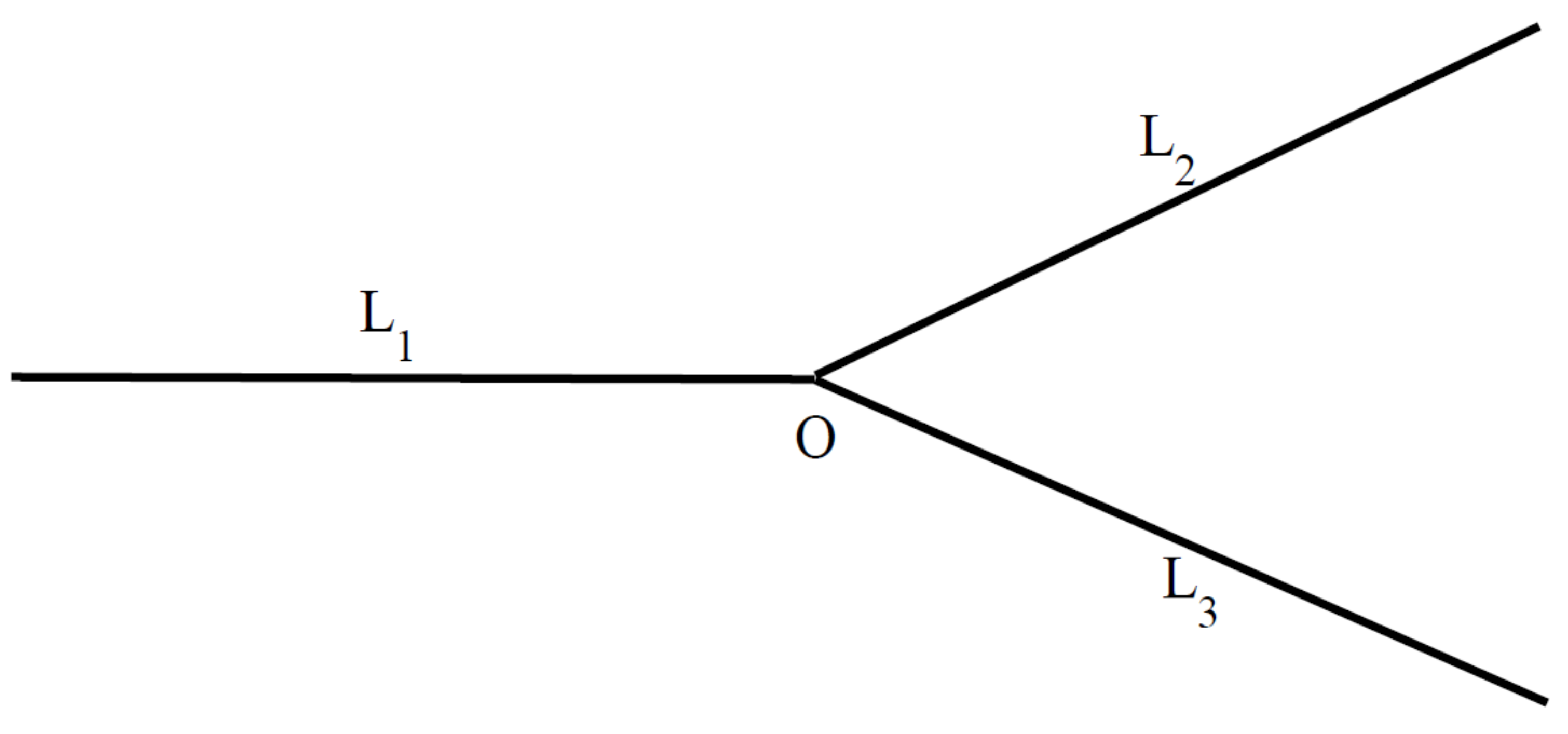}
\caption{Sketch of a metric  star graph. $L_j$ is the length of the $j$th bond with $j=1,2,3$.} \label{pic1}
\end{figure}
\section{Vertex boundary conditions and exact solutions for star graph}

The static sine-Gordon equation on a metric graph presented in Fig.\
\ref{pic1} can be written as \be
\frac{d^2}{dx^2}\phi_j=\frac{1}{\lambda_j^2}\sin(\phi_j),\,0 < x < L_j,
\label{ssge1}
\ee
where the wave functions $\phi_j$ are assigned to each bond of the
graph and  $j=1,2,3$ is the bond number. For wave equations on
networks, the connections of the network wires at the vertices
are provided by the vertex boundary conditions. In case of linear
wave equations, the underlying constraint to derive vertex boundary
conditions  is the self-adjointness of the problem
\ci{Kostrykin99, EK15}. However, for nonlinear case one should use
different conservation laws \ci{zar2010,adami2011,caputo14}. Here,
for the stationary sine-Gordon equation we impose the boundary conditions providing flux conservation at the vertex
 \be
\lambda_1\left.\frac{d\phi_1}{dx}\right|_{x=0}=\lambda_2\left.\frac{d\phi_2}{dx}\right|_{x=0}=\lambda_3\left.\frac{d\phi_3}{dx}\right|_{x=0}\label{ssgebc1}\ee
and the continuity of the weights of wave function derivatives, which are given as
\be
\lambda_1\phi_1|_{x=0}+\lambda_2\phi_2|_{x=0}+\lambda_3\phi_3|_{x=0}=0.\label{ssgebc2}
\ee
The boundary conditions at the end of each bond are imposed as
\be
\left.\frac{d\phi_j}{dx}\right|_{x=L_j}=2H_j. \label{ssgebc3} \ee
The boundary conditions given by Eqs.\ \re{ssgebc1}-\re{ssgebc3}
are consistent with other models of Josephson junction networks previously studied in \ci{Kogan2000,Hadi2005,Kupl2006,Kupl2007}. Exact solutions of
Eq.\ \re{ssge1} on a finite interval  have been obtained earlier in
\ci{Caputo2000,Hadi2005,Kupl2006,Kupl2007} for different special cases.
 Here, we  use an approach similar to that of the Refs.\ \ci{Kupl2006,Kupl2007} to obtain exact analytical solutions of Eq.\ \re{ssge1} for the boundary conditions \re{ssgebc1} and \re{ssgebc3}.

\subsection{Solution of type I}

Our purpose is to obtain exact analytical solutions of the problem given by Eqs.\
\re{ssge1}-\re{ssgebc3}.  A solution of Eq.\ (\ref{ssge1}) without boundary conditions can be written as \ci{Kupl2006,Kupl2007}
\be
\phi_j^{(\pm)}(x)=(2n_j+1)\pi\pm2\arcsin\left\{k_j\sn\left[\frac{x-x_{0,j}^{(\pm)}}{\lambda_j},k_j\right]\right\}\label{eq5}
\ee
where $k_j$ and $x_{0,j}^{(\pm)}$ are integration constants and $\sn$ is Jacobi's elliptic function. Depending on the value of $k_j$, the solution can be of two types. When $|H_j\lambda_j|\leq|k_j|\leq1$, we refer to the solution as solution of type 1 \ci{Kupl2006}. Taking into account that
\be
\frac{d\phi_j^{(\pm)}}{dx}=\pm\frac{2k_j}{\lambda_j}\cn\left[\frac{x-x_{0,j}^{(\pm)}}{\lambda_j},k_j\right],\label{eq6}
\ee
from boundary condition (\ref{ssgebc3}) we have
\be
x_{0,j}^{(\pm)}=L_j-\lambda_jF\left[\arccos\left(\pm\frac{H_j\lambda_j}{k_j}\right),k_j\right].\label{eq7}
\ee
Here, $\cn$ is Jacobi's elliptic function \ci{Yanke} and $F(\varphi,k)$ is the elliptic integral of the first kind \ci{Yanke}. Then solution of type 1 of the sine-Gordon equation on a metric star graph with the boundary conditions \re{ssgebc1}-\re{ssgebc3} can be written as
$$
\phi_j^{(\pm)}(x)= (2n_j+1)\pi\pm
$$
$$
\pm2\arcsin\left\{k_j\sn\left[\frac{x-L_j}{\lambda_j}
+F\left[\arccos\left(\pm\frac{H_j\lambda_j}{k_j}\right),k_j\right],k_j\right]\right\}
$$

The vertex boundary conditions  \re{ssgebc1} and \re{ssgebc2} lead to the following
system of transcendental equations for finding $k_j$:
\baa
\underset{j=1}{\overset{3}{\sum}}\lambda_j\arcsin\left\{k_j\sn\left[\frac{L_j}{\lambda_j}-F\left[\arccos\left(\pm\frac{H_j\lambda_j}{k_j}\right),k_j\right],k_j\right]\right\}\nonumber\\
=\pm\frac{1}{2}\underset{j=1}{\overset{3}{\sum}}(2n_j+1)\pi\lambda_j,\nonumber\\
\label{ste1}\eaa \baa
k_1\cn\left[\frac{L_1}{\lambda_1}-F\left[\arccos\left(\pm\frac{H_1\lambda_1}{k_1}\right),k_1\right],k_1\right]=\nonumber\\
=k_2\cn\left[\frac{L_2}{\lambda_2}-F\left[\arccos\left(\pm\frac{H_2\lambda_2}{k_2}\right),k_2\right],k_2\right],\label{ste2}\\
k_1\cn\left[\frac{L_1}{\lambda_1}-F\left[\arccos\left(\pm\frac{H_1\lambda_1}{k_1}\right),k_1\right],k_1\right]=\nonumber\\
=k_3\cn\left[\frac{L_3}{\lambda_3}-F\left[\arccos\left(\pm\frac{H_3\lambda_3}{k_3}\right),k_3\right],k_3\right].\label{ste3}
\eaa It is clear that if this system has roots, then our problem has solutions.
Here we obtain exact analytical solutions of this system  for two special cases.\\
\textit{Case I} is given by the relations
\baa
&\lambda_1=\lambda_2+\lambda_3,\,
\frac{L_j}{\lambda_j}=2mK(k_j),\;m\in{\bf N},\nonumber\\
&H_1\lambda_1=H_2\lambda_2=H_3\lambda_3=H>0,\nonumber\\
&n_1=-n,\,n_2=n_3=n-1,\,n\in{\bf Z}.\nonumber
\eaa
From Eqs.\ (\ref{ste1})-(\ref{ste3}) we have
$$
k_1=k_2=k_3=k
$$
and
$$
g^{(\pm)}(k)\equiv(-1)^{m+1}k\sqrt{1-\left(\frac{H}{k}\right)^2}=0,
$$ which gives
$$ k=\pm H.$$

\textit{Case II} corresponds to the constraints
\baa
&\lambda_1=\lambda_2+\lambda_3,\,
\frac{L_j}{\lambda_j}=(-1)^{m_j}p+2m_jK(k_j),\,m_j\in{N}\cup\{0\},\nonumber\\
&H_1\lambda_1=H_2\lambda_2=H_3\lambda_3=H>0,\nonumber\\
&n_1=-n,\,n_2=n_3=n-1,\,n\in{\bf Z},\nonumber \eaa where
$0<p<F\left[\arccos\left(H\right),1\right]$. Then from Eqs.\
\re{ste1}-\re{ste3} we have
$$
k_1=k_2=k_3=k.
$$
and
\be f^{(\pm)}(k)\equiv
p-F\left[\arccos\left(\pm\frac{H}{k}\right),k\right]=0.\label{t1c2k}
\ee

\begin{figure}[ht!]
\includegraphics[width=80mm]{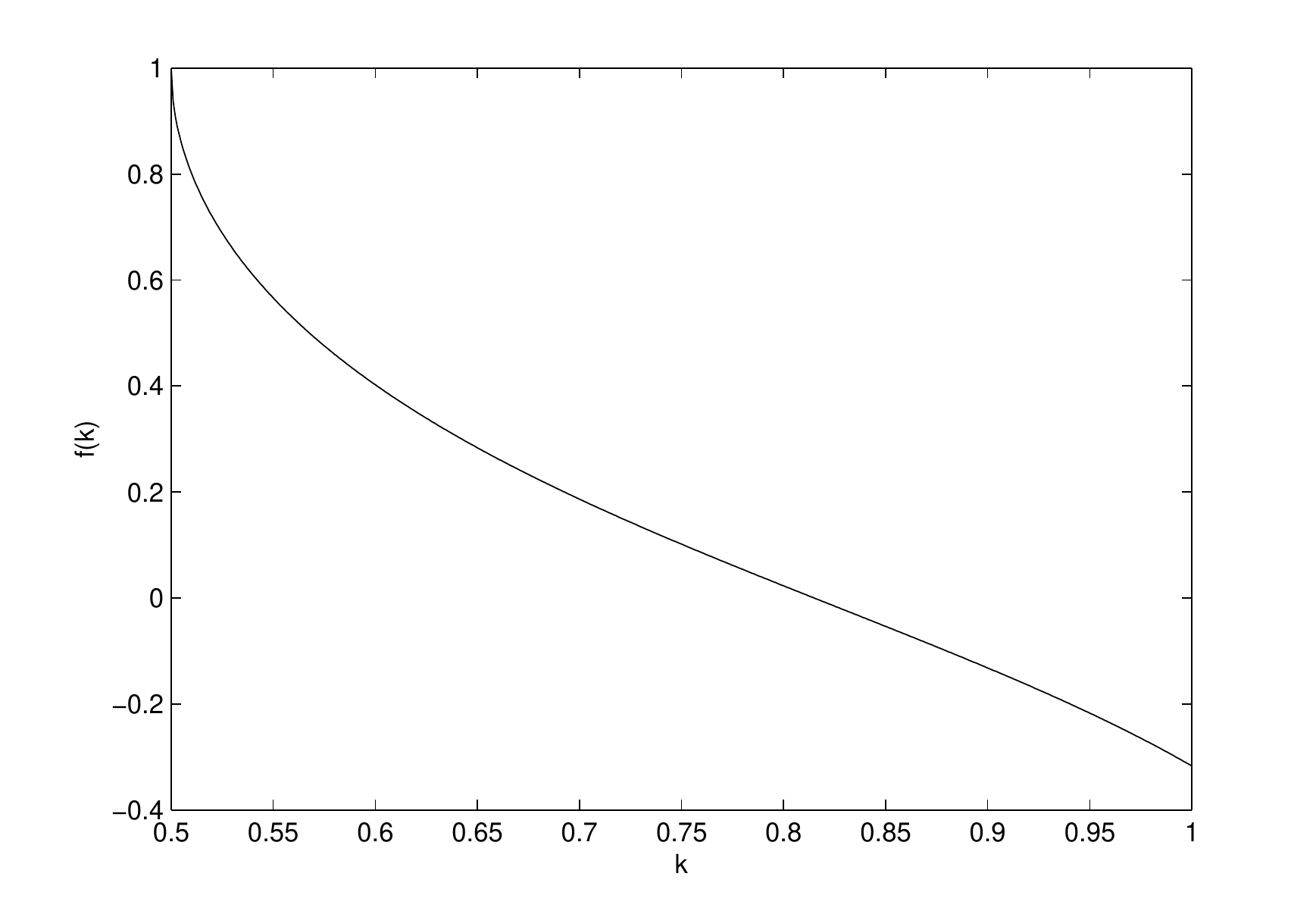}
\caption{Plot of the function $f(k)$ for
$H=0.5 \lambda_1=0.5, \lambda_2=0.2, \lambda_3=0.3,\;$ $m=1, p=1, n_1=-1, n_2=n_3=0,$
which shows the existence of a root of Eq.\ \re{t1c2k}} \label{root01}
\end{figure}

Since $f^{(\pm)}(\pm H)>0$, $f^{(\pm)}(\pm 1)<0$ and the functions
$f^{(+)}(k)$ and  ($f^{(-)}(k)$ are continuous on intervals  $[H;1]$ and $[-1;-H]$), respectively,
the system has at least one root. This can be seen from Fig.\ \ref{root01}  where the function $f(k)$ is plotted.

\subsection{Solutions of type II}

Solutions of type II for Eq.\ \re{ssge1}  are given by
\begin{equation} \phi^{(\pm)}_j(x)=\pi(2n_j+1)\pm
2\am\left(\frac{x-x^{(\pm)}_{0,j}}{\lambda_jk_j},k_j\right),\label{eq26}
\end{equation} and defined by the constraint
$$\frac{1}{\sqrt{1+H_j^2\lambda_j^2}}\leq|k_j|\leq\frac{1}{|H_j\lambda_j|}.$$
For the derivative of this solution, we have
\be\frac{d\phi_j^{(\pm)}(x)}{dx}=\pm\frac{2}{\lambda_jk_j}\dn\left(\frac{x-x^{(\pm)}_{0,j}}{\lambda_jk_j},k_j\right).\label{eq27}
\ee Inserting this derivative into the boundary condition \re{ssgebc3} we obtain \be
x_{0,j}^{(\pm)}=L_j\mp\lambda_jk_jF\left(\arcsin\frac{\sqrt{1-H_j^2\lambda_j^2k_j^2}}{k_j},k_j\right).\label{eq28}
\ee
Eqs.\ (\ref{eq26}) - (\ref{eq28}) together with the boundary conditions
\re{ssgebc1},\re{ssgebc2} lead to
\baa
\underset{j=1}{\overset{3}{\sum}}&\lambda_j\am\left[\frac{L_j}{\lambda_jk_j}\mp
F\left(\arcsin\frac{\sqrt{1-H_j^2\lambda_j^2k_j^2}}{k_j},k_j\right),k_j\right]=\nonumber\\
&\displaystyle=\pm\frac{1}{2}\underset{j=1}{\overset{3}{\sum}}(2n_j+1)\pi\lambda_j,\label{ste4}\\
&\frac{1}{k_1}\dn\left[\frac{L_1}{\lambda_1k_1}\mp
F\left(\arcsin\frac{\sqrt{1-H_1^2\lambda_1^2k_1^2}}{k_1},k_1\right),k_1\right]=\nonumber\\
&=\frac{1}{k_2}\dn\left[\frac{L_2}{\lambda_2k_2} \mp
F\left(\arcsin\frac{\sqrt{1-H_2^2\lambda_2^2k_2^2}}{k_2},k_2\right),k_2\right],\label{ste5}\\
&\frac{1}{k_1}\dn\left[\frac{L_1}{\lambda_1k_1}\mp F\left(\arcsin\frac{\sqrt{1-H_1^2\lambda_1^2k_1^2}}{k_1},k_1\right),k_1\right]=\nonumber\\
&\frac{1}{k_3}\dn\left[\frac{L_3}{\lambda_3k_3}\mp
F\left(\arcsin\frac{\sqrt{1-H_3^2\lambda_3^2k_3^2}}{k_3},k_3\right),k_3\right].\label{ste6}
\eaa

Again, one can show the existence of solutions of Eqs.\ \re{ste4}-\re{ste6} for two special cases. For \textit{case I}, which corresponds to the relations \baa
\underset{j=1}{\overset{3}{\sum}}\left(2n_j\mp2m+1\right)\lambda_j=0,\,m\in{\bf N}\nonumber\\
\frac{L_j}{\lambda_j}=2mk_jK(k_j),\;m\in{\bf N},\nonumber\\
H_1\lambda_1=H_2\lambda_2=H_3\lambda_3=H>1,
 \eaa
from Eqs.\ (\ref{ste4})-(\ref{ste6}), we have
\baa
&k_1=k_2=k_3=k,\nonumber\\
&g^{(\pm)}(k)\equiv
\am\left[F\left(\arcsin\frac{\sqrt{1-H^2k^2}}{k},k\right),k\right]=0.\label{eq36}
\eaa
Then, Eq.\ (\ref{eq36}) gives the following solution for the system of transcendental equations \re{ste4}-\re{ste6}:
$$
k=\pm \frac{1}{H}.
$$

For \textit{case II}, which is defined by the conditions
 \baa
\frac{L_j}{\lambda_j}=k_j\left(p+2m_jK(k_j)\right),\,m_j\in{N}\cup\{0\},\nonumber\\
H_1\lambda_1=H_2\lambda_2=H_3\lambda_3=H>1,\nonumber\\
\underset{j=1}{\overset{3}{\sum}}\left(2n_j\mp2m_j+1\right)\lambda_j=0,\nonumber
\eaa where $0<p<K\left(\frac{1}{\sqrt{1+H^2}}\right)$,  Eqs.\ (\ref{ste4})-(\ref{ste6}) yield
$$
k_1=k_2=k_3=k,
$$
that leads to \be f^{(\pm)}(k)\equiv p\mp
F\left[\arcsin\frac{\sqrt{1-H^2k^2}}{k},k\right]=0. \label{t2c2k}
\ee

\begin{figure}[ht!]
\includegraphics[width=80mm]{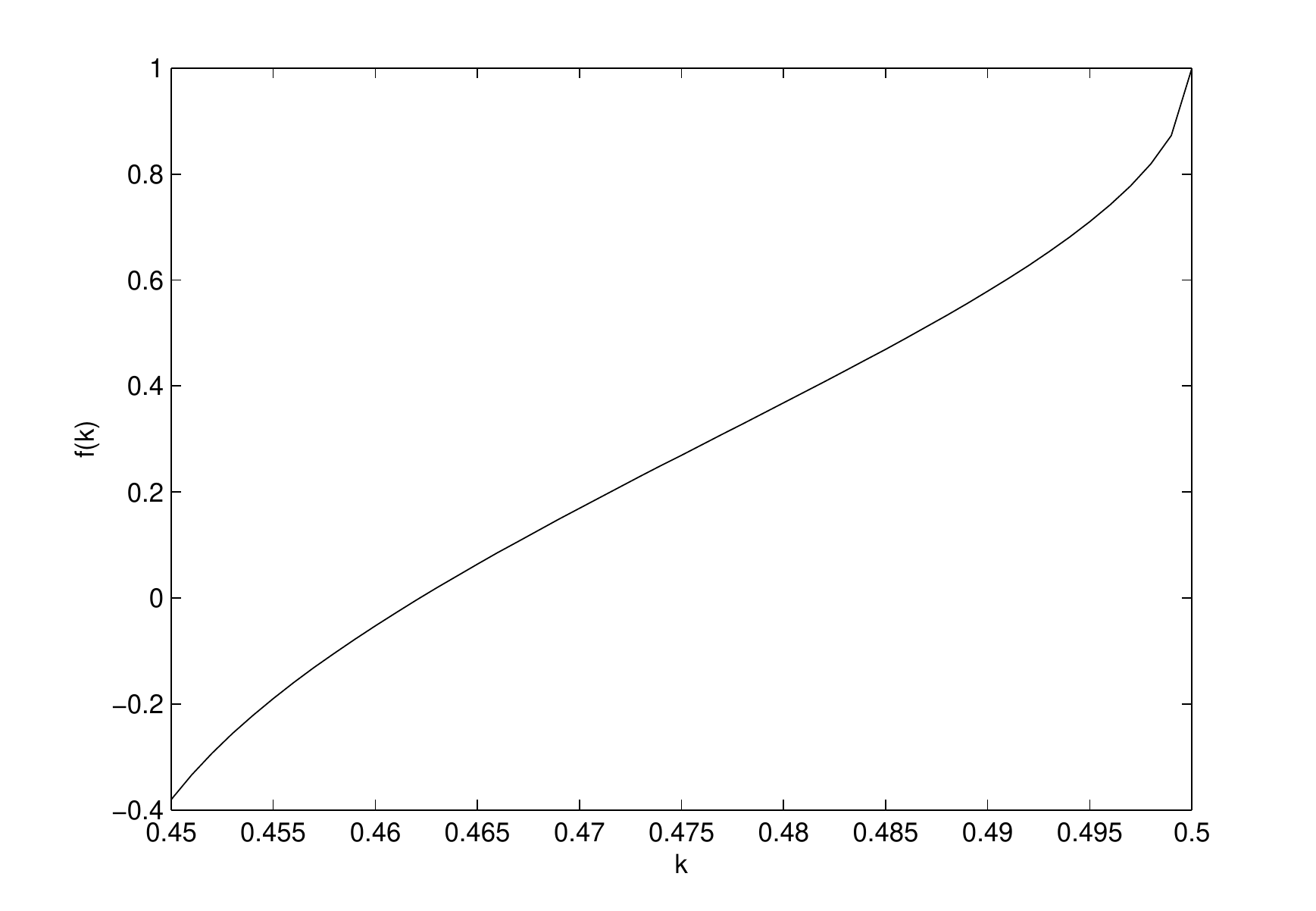}
\caption{ Plot of the function $f(k)$ for
$H=2 \lambda_1=0.5, \lambda_2=0.2, \lambda_3=0.3,\;$ $m_1=2,
m_2=m_3=1, p=1, n_1= n_2=n_3=1,$
 showing the existence of a root of  Eq.\ \re{t2c2k}}. \label{root02}
\end{figure}

Since $f^{(\pm)}(\pm \frac{1}{H})>0$, $f^{(\pm)}(\pm
\frac{1}{\sqrt{1+H^2}})<0$ and $f^{(+)}(k)$
$[\frac{1}{\sqrt{1+H^2}};\frac{1}{H}]$ ($f^{(-)}(k)$ is continuous
on interval $[-\frac{1}{H};-\frac{1}{\sqrt{1+H^2}}]$), it has at
least one root on this interval. Fig. \ref{root02} with the plot of $f(k)$
clearly shows that. 

\section{Applications of the method in tree graphs}

\begin{figure}[ht!]
\includegraphics[width=80mm]{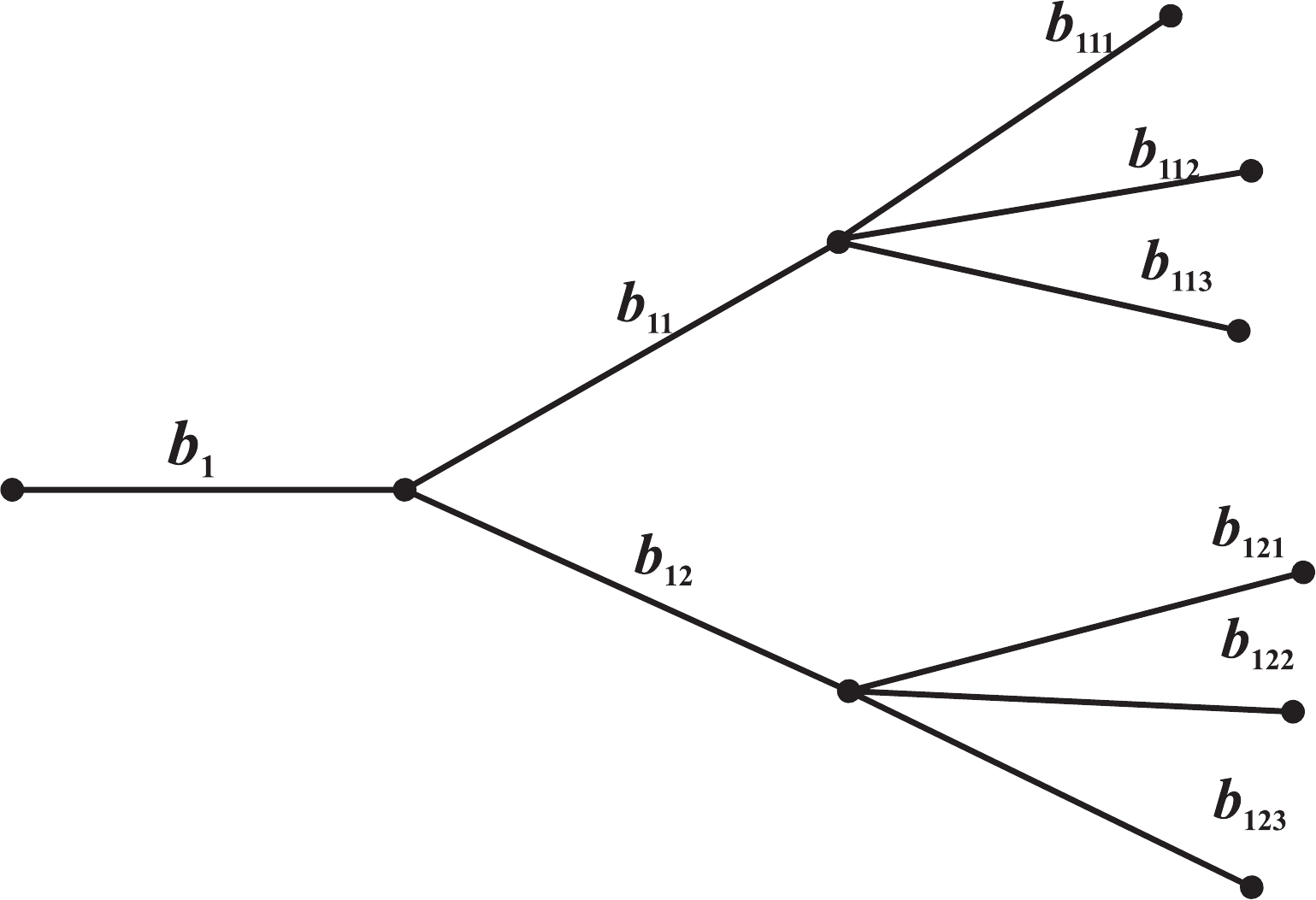}
\caption{A metric tree graph.} \label{tree1}
\end{figure}

The above discussion can also be applied to other simple topologies, such as tree graphs, loops and their combinations. Here, we briefly demonstrate this for the tree graph presented in Fig.\ \ref{tree1}.

The boundary conditions for each vertex and at the end of each bond
can be written as
\begin{eqnarray}
\left.\frac{d\phi_1}{dx}\right|_{x=0}=2H_1,\,\left.\frac{d\phi_{1ij}}{dx}\right|_{x=L_{1ij}}=2H_{1ij},\,i=1,2,\,j=1,2,3,\nonumber\\
\lambda_1\left.\frac{d\phi_1}{dx}\right|_{x=L_1}=\lambda_{1i}\left.\frac{d\phi_{1i}}{dx}\right|_{x=L_i},\,i=1,2,\nonumber\\
\lambda_{1i}\left.\frac{d\phi_{1i}}{dx}\right|_{x=L_{1i}}=\lambda_{1ij}\left.\frac{d\phi_{1ij}}{dx}\right|_{x=L_{1i}},\,i=1,2,\,j=1,2,3,\nonumber\\
\lambda_1\phi_1|_{x=L_1}+\lambda_{11}\phi_{11}|_{x=L_1}+\lambda_{12}\phi_{12}|_{x=L_1}=0,\nonumber\\
\lambda_{1i}\phi_{1i}|_{x=L_{1i}}+\underset{j=1}{\overset{3}{\sum}}\lambda_{1ij}\phi_{1ij}|_{x=L_{1i}}=0,\,i=1,2.\nonumber
\end{eqnarray}

Solutions of  type 1 and 2  of Eq.\ \re{ssge1} are defined similarly
to those for star graphs and can be written as

\baa
&\phi_b^{(\pm)}(x)=\left(2n_b+1\right)\pi\pm2\arcsin\left\{k_b\sn\left[\frac{x-x_{0,b}^{(\pm)}}{\lambda_b},k_b\right]\right\},\nonumber\\
&\phi_b^{(\pm)}(x)=\left(2n_b+1\right)\pi\pm2\am\left(\frac{x-x_{0,b}^{(\pm)}}{\lambda_bk_b},k_b\right).\nonumber
\eaa

Requiring these solutions to satisfy the boundary conditions leads
to a system of transcendental equations for finding $k_b$. Again,
exact solutions of this system can be obtained for two special
cases. However, unlike the case of star graphs, for tree graphs,
different bonds may have different type of solutions, e.g., one
subgraph can have a solution of type 1, while for others it is
possible to obtain the solution of type 2.

From the vertex boundary conditions we have the following system of
transcendental equations:
\baa
&\lambda_{1}\arcsin\left\{k_{1}\sn\left[\frac{L_1}{\lambda_1}-F\left[\arccos\left(\pm\frac{H_{1}\lambda_{1}}{k_{1}}\right),k_{1}\right],k_{1}\right]\right\}+\nonumber\\
&\lambda_{11}\arcsin\left\{k_{11}\sn\left[\frac{L_1-x_{0,11}^{(\pm)}}{\lambda_{11}},k_{11}\right]\right\}+\nonumber\\
&\lambda_{12}\arcsin\left\{k_{12}\sn\left[\frac{L_1-x_{0,12}^{(\pm)}}{\lambda_{12}},k_{12}\right]\right\}\nonumber\\
&=\mp\frac{\pi}{2}\left[(2n_{1}+1)\lambda_{1}+(2n_{11}+1)\lambda_{11}+(2n_{12}+1)\lambda_{12}\right],
\lab{trseq1}\\
&\lambda_{1i}\arcsin\left\{k_{1i}\sn\left[\frac{L_{1i}-x_{0,1i}^{(\pm)}}{\lambda_{1i}},k_{1i}\right]\right\}+\nonumber\\
&+\overset{3}{\underset{j=1}{\sum}}\lambda_{1ij}\arcsin\left\{k_{1ij}\sn\left[\frac{L_{1i}-L_{1ij}}{\lambda_{1ij}}\right.\right.+\nonumber\\
&+\left.\left.F\left[\arccos\left(\pm\frac{H_{1ij}\lambda_{1ij}}{k_{1ij}}\right),k_{1ij}\right],k_{1ij}\right]\right\}\nonumber\\
&=\mp\frac{\pi}{2}\left[(2n_{1i}+1)\lambda_{1i}+\overset{3}{\underset{j=1}{\sum}}(2n_{1ij}+1)\lambda_{1ij}\right],
\lab{trseq2}\\
&k_{1}\cn\left[\frac{L_1}{\lambda_1}-F\left[\arccos\left(\pm\frac{H_{1}\lambda_{1}}{k_{1}}\right),k_{1}\right],k_{1}\right]\nonumber\\
&=k_{1i}\cn\left[\frac{L_1-x_{0,1i}^{(\pm)}}{\lambda_{1i}},k_{1i}\right],
\lab{trseq3}\\
&k_{1i}\cn\left[\frac{L_1-x_{0,1i}^{(\pm)}}{\lambda_{1i}},k_{1i}\right]\nonumber\\
&=k_{1}\cn\left[\frac{L_{1i}-L_{1ij}}{\lambda_{1ij}}+F\left[\arccos\left(\pm\frac{H_{1ij}\lambda_{1ij}}{k_{1ij}}\right),k_{1ij}\right],k_{1ij}\right],
\lab{trseq4}
\eaa
where $i=1,2,\,j=1,2,3$. Choosing
$x_{0,1i}^{(\pm)}=\frac{1}{2}(L_{1i}+L_1)$ for case I we have
\baa
&\lambda_{1}+\lambda_{11}+\lambda_{12}=0,\,\lambda_{1i}+\overset{3}{\underset{j=1}{\sum}}\lambda_{1ij}=0,\nonumber\\
&\frac{L_{1}}{\lambda_{1}}=2mK(k_{1}),\,\frac{L_{1ij}-L_{1i}}{\lambda_{1ij}}=2mK(k_{1ij}),\nonumber\\
&\frac{L_{1i}-L_{1}}{2\lambda_{1i}}=2mK(k_{1i})-F\left[\arccos\left(\pm\frac{H_{1}\lambda_{1}}{k_{1i}}\right),k_{1i}\right],\nonumber\\
&H_1\lambda_1=H_{1ij}\lambda_{1ij}=H>0,\lab{heq}
\eaa
where $i=1,2,\,j=1,2,3,\,m\in{\bf Z}\setminus\{0\}$. Then simplifying the above system of transcendental equations \re{trseq1}-\re{trseq4} will yield
$$
k_b=k,\,g^{(\pm)}(k)\equiv(-1)^{m+1}k\sqrt{1-\left(\frac{H}{k}\right)^2}=0,
$$
which together with Eq.\ \re{heq} gives $k=\pm H$.

For case II we have \baa
&\lambda_{1}+\lambda_{11}+\lambda_{12}=0,\,\lambda_{1i}+\overset{3}{\underset{j=1}{\sum}}\lambda_{1ij}=0,\nonumber\\
&\frac{L_{1}}{\lambda_{1}}=(-1)^{m_1}p+2m_1K(k_{1}),\,\frac{L_{1ij}-L_{1i}}{\lambda_{1ij}}=(-1)^{m_{1ij}}p+2m_{1ij}K(k_{1ij}),\nonumber\\
&\frac{L_{1i}-L_{1}}{2\lambda_{1i}}=(-1)^{m_{1i}}p+2m_{1i}K(k_{1i})-F\left[arccos\left(\pm\frac{H_{1}\lambda_{1}}{k_{1i}}\right),k_{1i}\right],\nonumber\\&H_1\lambda_1=H_{1ij}\lambda_{1ij}=H>0,\nonumber
\eaa
where $0<p<F\left[arccos(H),1\right]$ and
$i=1,2,\,j=1,2,3,\,m\in{\bf Z}$. For this case the solution of Eqs.\ \re{trseq1}-\re{trseq4} can be written as
$$
k_b=k,\,f^{(\pm)}(k)\equiv
p-F\left[arccos\left(\pm\frac{H}{k}\right),k\right].
$$
Since $f^{(\pm)}(\pm \frac{1}{H})>0$, $f^{(\pm)}(\pm 1)<0$ and the function
$f^{(+)}(k)$  $[\frac{1}{H};1]$ ($f^{(-)}(k)$ is continuous on
interval $[-1;-\frac{1}{H}]$), it has at least one root on this
interval. We note that similarly, one can obtain solutions of mixed types.

\section{Stability of solutions}

Here we briefly analyze the stability of the obtained solutions using
the same method as in the Refs.\ci{Kupl2006,Kupl2007}. We do this
for a metric star graph, however, extending the method to tree graphs and
other graph topologies is trivial. First we define the  Gibbs
free-energy functional on the star graph presented in Fig.\ \ref{pic1}
\begin{equation}
\Omega_G=\underset{j=1}{\overset{3}{\sum}}\Omega_G^{(j)}\left[\phi_j,\,\frac{d\phi_j}{dx};\,H_j\right],\label{eq1}
\end{equation}
with the Gibbs free energy on each bond given by
\baa
&\Omega_G^{(j)}\left[\phi_j,\,\frac{d\phi_j}{dx};\,H_j\right]=2H_j^2W_j-2H_j\left[\phi_j(L_j)-\phi_j(0)\right]\nonumber\\
&+\frac{1}{\lambda_j}\underset{0}{\overset{L_j}{\int}}\left[1-cos\phi_j(x)+\frac{\lambda_j^2}{2}\left[\frac{d\phi_j(x)}{dx}\right]^2\right]dx.
\label{eq01} \eaa $W_j$ is the length of the bond $j$. It is easy to
see that the condition $\delta\Omega_G=0$ leads to the sine-Gordon equation on a star graph given by Eqs.\ \re{ssge1}-\re{ssgebc3}.

The key role in the stability analysis is played by the second variation
of the Gibbs functional given by
$$
\delta^2\Omega_G=\underset{j=1}{\overset{3}{\sum}}\frac{1}{\lambda_j}\underset{0}{\overset{L_j}{\int}}\left[cos\bar{\phi_j}(\delta\phi_j)^2+\lambda_j^2\left(\frac{d\delta\phi_j}{dx}\right)^2\right]dx.
$$
If for the tested solution of the sine-Gordon equation,  $\phi_j(x) =\bar \phi_j(x)$
$$
\delta^2\Omega_G\left[\phi_j,\frac{d\phi_j}{dx}\right]_{\phi_j(x)
=\bar\phi_j(x)} > 0,
$$
the solution will be inside the stability region  \ci{Kupl2006,Kupl2007}. For
$\delta^2\Omega_G\left[\phi_j,\frac{d\phi_j}{dx}\right]_{\phi_j(x)=\bar\phi_j(x)}$
having no definite sign, the solution will be unstable \ci{Kupl2006,Kupl2007}. The
condition
$$
\delta^2\Omega_G\left[\phi_j,\,\frac{d\phi_j}{dx}\right]_{\phi_j(x)
=\bar \phi_j(x)} \geq 0,
$$
defines the border of stability (bifurcation point). Furthermore,
following Refs.\ \ci{Kupl2006,Kupl2007},  these three conditions
can be reformulated in terms of  $\mu_0$, the lowest eigenvalue
of the Sturm-Liouville eigenvalue problem
\begin{eqnarray}
-\lambda_j^2\frac{d^2\psi_j}{dx^2}+cos\bar{\phi_j}\psi_j=\mu\psi_j,\label{eq12}\\
\lambda_1\left.\frac{d\psi_1}{dx}\right|_{x=0}=\lambda_2\left.\frac{d\psi_2}{dx}\right|_{x=0}=\lambda_3\left.\frac{d\psi_3}{dx}\right|_{x=0}\label{bc1}\\
\lambda_1\psi_1|_{x=0}+\lambda_2\psi_2|_{x=0}+\lambda_3\psi_3|_{x=0}=0,\label{bc2}\\
\left.\frac{d\psi_j}{dx}\right|_{x=L_j}=0,\,j=1,2,3.\label{bc3}
\end{eqnarray}
If the lowest eigenvalue $\mu=\mu_0$ of this Sturm-Liouville problem is negative, i.e.\ $\mu_0<0$, the solution $\phi =\phi(y)$ corresponds to a saddle point of Eq.\ \re{eq01} and therefore is unstable. The stable solutions minimize the
functional $\Omega_G$ and are characterized by $\mu_0>0$. The
boundary between  stable and unstable solutions are determined by
the condition $\mu=0$. By solving numerically the problem \re{eq12} -\re{bc3} we found that $\mu_0<0$ for both cases of the solutions of type 1.
For the  case I of the solution of type 2 we have $\mu_0>0$, while for the case II
of the solution of type 2 we found that $\mu_0<0$.
Therefore only case I of the solution of type 2 is stable, while the other solutions are unstable.

\section{Conclusions}
In this paper, we have studied the stationary sine-Gordon equation on
simple metric graphs by imposing the vertex boundary conditions following from the flux conservation and the continuity of the  weights of the wave function derivatives. Exact analytical solutions are obtained for a metric star graph.
The constraints allowing such exact  solutions are determined in
terms of bond nonlinearity coefficients.

The treatment has been extended to metric tree graphs and
explicit solutions are derived. Generalizations to other simple
topologies such as loop graphs and combinations of loop and star
graphs have also been discussed. The stability of the obtained solutions has been
analyzed. The obtained results can be directly applied to the
study of static solitons in 2D branched  Josephson junctions in a constant magnetic field, i.e.\ T-,  Y- and tree-shaped versions of the model studied in \ci{Kupl2006}. 
Finally, we note that the method can be extended to the case of "current carrying" boundary conditions studied in Ref.\ \ci{Kupl2007}.

\section{Acknowledgements}

We thank Dmitry Pelinovsky for his useful comments on the paper.
This work is supported by a grant of the Volkswagen Foundation.
The work of KS is partially supported by the grant of the Committee for the Coordination Science and Technology Development (Ref.Nr. F-2-003).

\bibliography{biblar}
\bibliographystyle{apsrev4-1}

\end{document}